



\font\twelverm=cmr10 scaled 1200    \font\twelvei=cmmi10 scaled 1200
\font\twelvesy=cmsy10 scaled 1200   \font\twelveex=cmex10 scaled 1200
\font\twelvebf=cmbx10 scaled 1200   \font\twelvesl=cmsl10 scaled 1200
\font\twelvett=cmtt10 scaled 1200   \font\twelveit=cmti10 scaled 1200

\skewchar\twelvei='177   \skewchar\twelvesy='60


\def\twelvepoint{\normalbaselineskip=12.4pt
  \abovedisplayskip 12.4pt plus 3pt minus 9pt
  \belowdisplayskip 12.4pt plus 3pt minus 9pt
  \abovedisplayshortskip 0pt plus 3pt
  \belowdisplayshortskip 7.2pt plus 3pt minus 4pt
  \smallskipamount=3.6pt plus1.2pt minus1.2pt
  \medskipamount=7.2pt plus2.4pt minus2.4pt
  \bigskipamount=14.4pt plus4.8pt minus4.8pt
  \def\rm{\fam0\twelverm}          \def\it{\fam\itfam\twelveit}%
  \def\sl{\fam\slfam\twelvesl}     \def\bf{\fam\bffam\twelvebf}%
  \def\mit{\fam 1}                 \def\cal{\fam 2}%
  \def\tt{\twelvett}
  \textfont0=\twelverm   \scriptfont0=\tenrm   \scriptscriptfont0=\sevenrm
  \textfont1=\twelvei    \scriptfont1=\teni    \scriptscriptfont1=\seveni
  \textfont2=\twelvesy   \scriptfont2=\tensy   \scriptscriptfont2=\sevensy
  \textfont3=\twelveex   \scriptfont3=\twelveex  \scriptscriptfont3=\twelveex
  \textfont\itfam=\twelveit
  \textfont\slfam=\twelvesl
  \textfont\bffam=\twelvebf \scriptfont\bffam=\tenbf
  \scriptscriptfont\bffam=\sevenbf
  \normalbaselines\rm}



\def\beginlinemode{\endmode
  \begingroup\parskip=0pt \obeylines\def\\{\par}\def\endmode{\par\endgroup}}
\def\beginparmode{\endmode
  \begingroup \def\endmode{\par\endgroup}}
\let\endmode=\par
{\obeylines\gdef\
{}}
\def\singlespace{\baselineskip=\normalbaselineskip}
\def\oneandathirdspace{\baselineskip=\normalbaselineskip
  \multiply\baselineskip by 4 \divide\baselineskip by 3}
\def\oneandahalfspace{\baselineskip=\normalbaselineskip
  \multiply\baselineskip by 3 \divide\baselineskip by 2}
\def\doublespace{\baselineskip=\normalbaselineskip \multiply\baselineskip by 2}

\newcount\firstpageno
\firstpageno=2
\footline={\ifnum\pageno<\firstpageno{\hfil}\else{\hfil\twelverm\folio\hfil}\fi}
\let\rawfootnote=\footnote		
\def\footnote#1#2{{\rm\singlespace\parindent=0pt\rawfootnote{#1}{#2}}}
\def\raggedcenter{\leftskip=4em plus 12em \rightskip=\leftskip
  \parindent=0pt \parfillskip=0pt \spaceskip=.3333em \xspaceskip=.5em
  \pretolerance=9999 \tolerance=9999
  \hyphenpenalty=9999 \exhyphenpenalty=9999 }
\def\dateline{\rightline{\ifcase\month\or
  January\or February\or March\or April\or May\or June\or
  July\or August\or September\or October\or November\or December\fi
  \space\number\year}}
\def\received{\vskip 3pt plus 0.2fill
 \centerline{\sl (Received\space\ifcase\month\or
  January\or February\or March\or April\or May\or June\or
  July\or August\or September\or October\or November\or December\fi
  \qquad, \number\year)}}


\hsize=6.5truein
\hoffset=0truein
\vsize=8.9truein
\voffset=0truein
\parskip=\medskipamount
\twelvepoint		
\doublespace		
\overfullrule=0pt	


\def\preprintno#1{
 \rightline{\rm #1}}	

\def\title			
  {\null\vskip 3pt plus 0.2fill
   \beginlinemode \doublespace \raggedcenter \bf}

\def\author			
  {\vskip 3pt plus 0.2fill \beginlinemode
   \singlespace \raggedcenter}

\def\affil			
  {\vskip 3pt plus 0.1fill \beginlinemode
   \oneandahalfspace \raggedcenter \sl}

\def\abstract			
  {\vskip 3pt plus 0.3fill \beginparmode
   \doublespace \narrower ABSTRACT: }

\def\endtitlepage		
  {\endpage			
   \body}

\def\body			
  {\beginparmode}		

\def\head#1{			
  \filbreak\vskip 0.5truein	
  {\immediate\write16{#1}
   \raggedcenter \uppercase{#1}\par}
   \nobreak\vskip 0.25truein\nobreak}

\def\subhead#1{			
  \vskip 0.25truein		
  {\raggedcenter #1 \par}
   \nobreak\vskip 0.25truein\nobreak}

\def\refto#1{$^{#1}$}		

\def\references			
  {\head{References}		
   \beginparmode
   \frenchspacing \parindent=0pt \leftskip=1truecm
   \parskip=8pt plus 3pt \everypar{\hangindent=\parindent}}

\gdef\refis#1{\indent\hbox to 0pt{\hss#1.~}}	

\gdef\journal#1, #2, #3, 1#4#5#6{		
    {\sl #1~}{\bf #2}, #3 (1#4#5#6)}		

\gdef\journ2 #1, #2, #3, 1#4#5#6{		
    {\sl #1~}{\bf #2}: #3 (1#4#5#6)}		

\def\refstylenp{		
  \gdef\refto##1{ [##1]}				
  \gdef\refis##1{\indent\hbox to 0pt{\hss##1)~}}	
  \gdef\journal##1, ##2, ##3, ##4 {			
     {\sl ##1~}{\bf ##2~}(##3) ##4 }}

\def\refstyleprnp{		
  \gdef\refto##1{ [##1]}				
  \gdef\refis##1{\indent\hbox to 0pt{\hss##1)~}}	
  \gdef\journal##1, ##2, ##3, 1##4##5##6{		
    {\sl ##1~}{\bf ##2~}(1##4##5##6) ##3}}

\def\prd{\journal Phys. Rev. D, }

\def\prl{\journal Phys. Rev. Lett., }

\def\endreferences{\body}

\def\figurecaptions		
  {\endpage
   \beginparmode
   \head{Figure Captions}
}

\def\endfigurecaptions{\body}

\def\endpage			
  {\vfill\eject}

\def\endpaper			
  {\endmode\vfill\supereject}

\def\endit
  {\endpaper\end}


\def\ref#1{Ref. #1}			
\def\Ref#1{Ref. #1}			

\def\frac#1#2{{\textstyle #1 \over \textstyle #2}}

\def\sla{\raise.15ex\hbox{$/$}\kern-.57em}
\def\leaderfill{\leaders\hbox to 1em{\hss.\hss}\hfill}
\def\twiddle{\lower.9ex\rlap{$\kern-.1em\scriptstyle\sim$}}
\def\bigtwiddle{\lower1.ex\rlap{$\sim$}}
\def\gtwid{\mathrel{\raise.3ex\hbox{$>$\kern-.75em\lower1ex\hbox{$\sim$}}}}
\def\ltwid{\mathrel{\raise.3ex\hbox{$<$\kern-.75em\lower1ex\hbox{$\sim$}}}}
\def\square{\kern1pt\vbox{\hrule height 1.2pt\hbox{\vrule width 1.2pt\hskip 3pt
   \vbox{\vskip 6pt}\hskip 3pt\vrule width 0.6pt}\hrule height 0.6pt}\kern1pt}

\def\caption#1{\centerline{
	\uppercase{#1}}\noindent}
\newdimen\psfigsize
\newcount\psfigcount
\def\psfigure#1 #2 #3 #4 #5{\topinsert\vbox{
    \psfigcount=#1
    \psfigsize=#1truept
    \vskip \psfigsize
    \includegraphics{#4}
    \vskip 10truept
    \caption {#3}
    {\narrower\singlespace\noindent#5\par}\vskip 0.1truein
    plus0.2truein}
\endinsert}
%
%
%
\def\psoddfigure#1 #2 #3 #4 #5 #6{\topinsert\vbox{
    \psfigcount=#1
    \psfigsize=#2truept
    \vskip \psfigcount truept
    \includegraphics{#5}
    \advance\psfigsize by -\psfigcount truept \vskip\psfigsize
    \vskip 10truept
    \caption {#4}
    {\narrower\singlespace\noindent#6\par}\vskip 0.1truein
    plus0.2truein}
\endinsert}
%
\def\figurespace#1 #2 #3 #4 {\topinsert\vbox{
    \psfigcount=#1
    \psfigsize=#1truept
    \vskip \psfigsize
    \vskip 10truept
    \caption {#3}
    {\narrower\singlespace\noindent#4\par}\vskip 0.1truein
    plus0.2truein}
\endinsert}

\def\preprintno#1{
 \rightline{\rm #1}}	

\def\TABLEcap#1#2{
 {\narrower\medskip\singlespace\noindent{ Table~#1.}#2 \smallskip
   \par
}}

\def\dbline{\noalign{\hrule}\noalign{\vskip 2pt}\noalign{\hrule}}
\def\sgline{\noalign{\hrule}}
\def\notext{\omit&\omit&\omit&\omit&\omit&\omit\cr}

\def\fstrut{\vrule height 13.5pt depth 2.25pt width 0pt}

\def\endrule{&\omit\fstrut\vrule\cr}


\def\Tr{\rm Tr}

\catcode`@=11
\newcount\r@fcount \r@fcount=0
\newcount\r@fcurr
\immediate\newwrite\reffile
\newif\ifr@ffile\r@ffilefalse
\def\w@rnwrite#1{\ifr@ffile\immediate\write\reffile{#1}\fi\message{#1}}

\def\writer@f#1>>{}
\def\referencefile{
  \r@ffiletrue\immediate\openout\reffile=\jobname.ref%
  \def\writer@f##1>>{\ifr@ffile\immediate\write\reffile%
    {\noexpand\refis{##1} = \csname r@fnum##1\endcsname = %
     \expandafter\expandafter\expandafter\strip@t\expandafter%
     \meaning\csname r@ftext\csname r@fnum##1\endcsname\endcsname}\fi}%
  \def\strip@t##1>>{}}

\def\citeall#1{\xdef#1##1{#1{\noexpand\cite{##1}}}}
\def\cite#1{\each@rg\citer@nge{#1}}	

\def\each@rg#1#2{{\let\thecsname=#1\expandafter\first@rg#2,\end,}}
\def\first@rg#1,{\thecsname{#1}\apply@rg}	
\def\apply@rg#1,{\ifx\end#1\let\next=\relax
\else,\thecsname{#1}\let\next=\apply@rg\fi\next}

\def\citer@nge#1{\citedor@nge#1-\end-}	
\def\citer@ngeat#1\end-{#1}
\def\citedor@nge#1-#2-{\ifx\end#2\r@featspace#1 
  \else\citel@@p{#1}{#2}\citer@ngeat\fi}	
\def\citel@@p#1#2{\ifnum#1>#2{\errmessage{Reference range #1-#2\space is bad.}%
    \errhelp{If you cite a series of references by the notation M-N, then M and
    N must be integers, and N must be greater than or equal to M.}}\else%
 {\count0=#1\count1=#2\advance\count1
by1\relax\expandafter\r@fcite\the\count0,%
  \loop\advance\count0 by1\relax
    \ifnum\count0<\count1,\expandafter\r@fcite\the\count0,%
  \repeat}\fi}

\def\r@featspace#1#2 {\r@fcite#1#2,}	
\def\r@fcite#1,{\ifuncit@d{#1}
    \newr@f{#1}%
    \expandafter\gdef\csname r@ftext\number\r@fcount\endcsname%
                     {\message{Reference #1 to be supplied.}%
                      \writer@f#1>>#1 to be supplied.\par}%
 \fi%
 \csname r@fnum#1\endcsname}
\def\ifuncit@d#1{\expandafter\ifx\csname r@fnum#1\endcsname\relax}%
\def\newr@f#1{\global\advance\r@fcount by1%
    \expandafter\xdef\csname r@fnum#1\endcsname{\number\r@fcount}}

\let\r@fis=\refis			
\def\refis#1#2#3\par{\ifuncit@d{#1}
   \newr@f{#1}%
   \w@rnwrite{Reference #1=\number\r@fcount\space is not cited up to now.}\fi%
  \expandafter\gdef\csname r@ftext\csname r@fnum#1\endcsname\endcsname%
  {\writer@f#1>>#2#3\par}}

\def\ignoreuncited{
   \def\refis##1##2##3\par{\ifuncit@d{##1}%
     \else\expandafter\gdef\csname r@ftext\csname
r@fnum##1\endcsname\endcsname%
     {\writer@f##1>>##2##3\par}\fi}}

\def\r@ferr{\endreferences\errmessage{I was expecting to see
\noexpand\endreferences before now;  I have inserted it here.}}
\let\r@ferences=\references
\def\references{\r@ferences\def\endmode{\r@ferr\par\endgroup}}

\let\endr@ferences=\endreferences
\def\endreferences{\r@fcurr=0
  {\loop\ifnum\r@fcurr<\r@fcount
    \advance\r@fcurr by 1\relax\expandafter\r@fis\expandafter{\number\r@fcurr}%
    \csname r@ftext\number\r@fcurr\endcsname%
  \repeat}\gdef\r@ferr{}\endr@ferences}


\let\r@fend=\endpaper\gdef\endpaper{\ifr@ffile
\immediate\write16{Cross References written on []\jobname.REF.}\fi\r@fend}

\catcode`@=12

\citeall\refto		
\citeall\ref		%
\citeall\Ref		%

\def\preprint{N}


\newcount\notenumber
\notenumber=0
\def\note#1{\advance\notenumber by 1\parindent=0pt\baselineskip=12pt
   \footnote{$^{\the\notenumber}$}{\noindent #1\par}}
\newcount\section
\newcount\subsection
\section=0 \subsection=0
\def\advancesct{\global\advance\section by1\global\subsection=0\relax}
\def\advancesubsct{\global\advance\subsection by 1 }
\def\heading#1{\par
   \noindent{\bf\advancesct\number\section. #1}\par}

\headline{\hfill}
\overfullrule=0pt

\def\1o2{{1 \over 2}}

\def\b{\beta}
\def\D{\Delta}

\def\part{\partial}

\def\1bar{1{\hskip-2pt\hbox{\vrule height7pt width.75pt}}}
\def\ls{\hbox{\raise.6ex\hbox{$<$}\llap{\lower.4ex\hbox{$\sim$}}}}
\null
\if \preprint Y \twelvepoint\oneandathirdspace \fi
\vbox to 3.1truein {
\title
Heavy Dynamical Fermions in Lattice QCD
\author
Anna Hasenfratz
\author
Thomas A. DeGrand
\affil
Department of Physics
University of Colorado
Boulder, Colorado 80309
\endgroup
\vfill
}
\if \preprint Y \preprintno{COLO-HEP-311}\fi
\abstract
It is expected that the only effect of heavy dynamical fermions in QCD is
to renormalize the gauge coupling. We derive a simple expression for
the shift in the gauge coupling induced by $N_f$ flavors of heavy fermions.
We compare this formula to the shift in the gauge coupling  at
which the confinement-deconfinement phase transition  occurs (at fixed
lattice size) from numerical simulations
 as a function of quark mass and $N_f$. We  find
 remarkable agreement with our expression down to a  fairly light quark mass.
However, simulations with eight heavy flavors and two light flavors show
that the eight flavors do more than just shift the gauge coupling.
We observe confinement-deconfinement transitions at $\beta=0$ induced by
a large number of heavy quarks. We comment on the relevance of our results to
contemporary simulations of QCD which include dynamical fermions.
\endtitlepage
\heading{Introduction}
QCD investigations frequently deal with the effect of heavy fermions
either as real physical effects (heavy quarks) or as the consequence
of the regularization (Wilson fermions). In all cases
the influence of heavy fermions at low energies was expected to be no more
than some induced effective gauge coupling.

Finite temperature
simulations with two light and one heavier quarks do not show a
significant difference from two light quark simulations. The effect of
the heavy fermion can be described to a good approximation by a
shift of the gauge coupling, $\D\b\approx 0.08$ for $m_q=0.25$
(Ref. \cite{KS2+1}) and $\D\b\approx 0.13$ for $m_q=0.1$ (Ref.
\cite{COLUM2+1}).

Wilson fermions have 15 heavy fermion doublers for each light fermion
yet the spectrum hardly differs from  the spectrum of staggered fermions
if one takes into account the doublers by shifting the gauge coupling.
For two flavors at a gauge coupling around $\b=5.6$
and hopping parameter value $\kappa\approx 0.16$ this  shift is about
$\D\b\approx 0.3$ - the only apparent effect of the doublers.

When can we expect that the fermions influence the physical spectrum
in a non-trivial way and when can we just replace them with an
effective local gauge action?
The answer obviously depends on the physical processes we are
investigating.  Heavy fermions are always present in
the spectrum, unless their mass is above the cut-off, but if the low lying
gauge and light quark hadronic spectrum is much below the energy level of
the heavy fermions they will not directly influence the low energy spectrum.

The fermions' induced gauge coupling can be calculated by evaluating a 1-loop
graph if the fermions are heavy. This analysis was presented in Ref. \cite{HH}
using dimensional regularization,  where  the possibility of generating a
continuum gauge theory with heavy fermions was investigated.
The lattice regularized calculation is
briefly mentioned in Ref. \cite{L92}.
In this paper we analyze further the  analytically predicted  induced gauge
coupling and compare it to existing and new numerical
results.

\heading{The induced gauge coupling on the lattice}

Consider the lattice regularized model of $\tilde {N_f}$ fundamental
(Wilson) fermions
interacting with SU(3) gauge fields, whose action is
$$
S=\b\sum_{n,\mu} Tr(U_p)+
{1\over 2\kappa}\sum_{n,m} {\bar\psi}_n K_{nm}[U]\psi_m,
\eqno{(1)}
$$
where
$$
K_{nm}[U]=\delta_{nm}-\kappa\sum_\mu((r-\gamma_\mu)U_{n\mu}\delta_{n+\mu,m}
+(r+\gamma_\mu)U^{\dagger}_{n\mu}\delta_{n-\mu,m}).
\eqno{(2)}
$$
$\kappa$ is related to the inverse of the bare fermion mass
$$
\kappa= {1\over 2ma+8r},
\eqno{(3)}
$$
where $a$ is the dimensional lattice spacing.
$r=1$ corresponds to the usual Wilson fermion formulation while $r=0$
describes $N_f=16 \times\tilde{N_f}$   staggered fermions.
Integrating out the fermions we obtain the effective gauge action
$$
\eqalign{
S_{eff}&=S_g- \Tr \ln K[U] \cr
&=S_g+\sum_{\Gamma}\kappa^{l[\Gamma]}{1\over l[\Gamma]}
\Tr\left (\prod_\Gamma(r\pm\gamma_\mu)\right )\cdot
\left (\Tr U[\Gamma]+\Tr U^{\dagger}[\Gamma]\right ),
}\eqno{(4)}
$$
where the sum is over all closed gauge loops $\Gamma$ and $l[\Gamma]$ is
the length of the loop.
Using the continuum representation of
 the gauge field  $U_{n\mu}=e^{iagA_\mu(n)}$ one can express
$S_{eff}^{ferm}$ in terms of the continuum fields $A_{\mu}(n)$ as the
sum of one loop diagrams
\if \preprint N \vfill\eject \fi
$$
S_{eff}^{ferm}=figure
\eqno{(5)}
$$
\vskip 2in
The leading term of the effective action is the usual continuum gauge
action ${1 \over g^2_0} F_{\mu\nu}F_{\mu\nu}$ where the coefficient
${1/ g^2_0}$ can be
calculated by evaluating the two 2-legged graphs in Eqn. 5.
The quantity  $1 / g^2_0$ can also be calculated starting
with  Eqn. 4 and using the method presented in Ref. \cite{MK} for
adjoint scalars. It happens that
this technique is  actually incorrect for adjoint fields but
correct for fundamental ones.

The result is given
by a four-dimensional lattice integral
$$
{1 \over g^2_{0}}=
{\tilde{N_f} \over 4} \int {{d^4p }\over {(2\pi)^4}}\Tr\left \{
Q(p_\mu)S(p)Q(p_\mu){\partial^2 \over \partial p^2_\nu} S(p)\right \}
\eqno{(6)}
$$
where $S(p)$ is the lattice fermion propagator
$$
S^{-1}(p)=\frac{1}{2\kappa}-r\sum_\mu \cos(p_\mu)-i\sum_\mu \gamma_\mu
\sin(p_\mu)
\eqno{(7)}
$$
and $Q(p)$ is given by
$$
Q(p_\mu)=i r \sin(p_\mu) +\gamma_\mu \cos(p_\mu).
\eqno{(8)}
$$
The integral reduces to the hopping parameter expansion result in the
$\kappa \to 0$ limit
$$\eqalign{
& {1 \over g^2_{0}}=4\tilde{N_f} \kappa^4 ,\quad r=1 ,\cr
& {1 \over g^2_{0}}=2\tilde{N_f} \kappa^4 ,\quad r=0 .
}\eqno{(9)}
$$
For small $ma$ ($\kappa \to 0.125$) it has a logarithmic singularity
$$
{1 \over g^2_{0}}=16 {\tilde {N_f}\over {24\pi^2}} \ln {\pi^2\over
{m^2a^2}},\quad r=0 .
\eqno{(10)}
$$
The effective action has additional terms containing more derivatives
and/or external gluon legs. These graphs are multiplied by negative
powers of $m$ and are suppressed for heavy fermions\refto{HH}.
In the limit where
the higher order terms can be neglected the effective action is indeed a
pure gauge action with bare coupling constant given by Eqn. 6. In terms of
the plaquette action lattice model it corresponds to  an effective
plaquette term with coefficient $\D\b=6/g_0^2$. Table 1 shows $\D\b$ for
several $m$ values for $r=0$ and $N_f=16\tilde{N_f}=1$ fermion flavor.

\heading{Validity of the effective action}

In this section we investigate under what conditions a single plaquette
effective action can describe the fermionic theory at low energies.

The question is two-fold: 1) Can the non-local effective action
Eqn. 4 indeed be replaced by a single plaquette
 term and 2) how well does Eqn. 6
predict the coefficient of this term? It is possible to
have a pure gauge effective action in a region where Eqn. 6 is no longer
 valid. We will consider the second point in the next chapter and now
investigate the first question.

The low energy effective theory can be considered pure gauge if the
gluonic spectrum characterized by the $\Lambda$ parameter is much lower
than the fermionic mass scale that is characterized by the fermion mass
$m$.

The one-loop definition of the lattice $\Lambda$ parameter is
$$
         \Lambda_{latt} = {1 \over a}
         \exp \left \{ - {\b_{eff} \over {12 \beta_0}
          } \right \}.
\eqno{(11)}
$$
where $\beta_0=11N_c/48\pi^2$ is the first (universal) coefficient of the
$\beta$-function and $\b_{eff}=\b+\D\b$.
The condition $\Lambda_{latt}<<m$ can be expressed as
$$
{{\b+\D\b}\over {12\b_0}} +\ln(am) >>0.
\eqno{(12)}
$$
As $\D\b$ is proportional to $N_f$, this condition can be translated into
a lower limit on the fermion flavors. For example, for $\b=5.7$, $ma=0.1$,
assuming the validity of Eqn. 6, if $N_f >>-28$ (i.e. for any physical
$N_f$)  the heavy fermions and
the gluonic sector decouple.

It is interesting to consider the $\b=0$ strong gauge coupling limit.
For small $m$ $\D\b$ is logarithmically divergent leading to the condition
for decoupling
$$N_f  > {33 \over 2} \eqno{(13)} $$
in the $ a \to 0$ limit (Recall $N_f = 16 \tilde N_f)$.
The minimum number of flavors coincides with the value where the $\beta$
function of the gauge-fermion system changes sign and becomes
non-asymptotically free. Since the presence of a gauge term ($\b\ne 0$)
relaxes the limit on $N_f$,  if Eqn. 6 holds in an SU(3) gauge theory with
17 or more fermions, then the fermions always decouple from
the low lying gluonic
spectrum. An identical condition was found in Ref. \cite{HH} using dimensional
regularization. An interesting consequence is that the $m=0$ theory is
always deconfined even in the strong gauge coupling limit
for $N_f\ge17$ flavors,
assuming $\D\b$ diverges as in Eqn. 10.
One should keep in mind, however, that
this derivation is valid only if the
higher order terms in the effective action can be neglected.

\heading{Examples}
Now we want to address the second question. When a
 gauge-fermion theory with massive fermions can be replaced by a gauge
 theory with a single plaquette action, is the shift in $\beta$
induced by the fermions  given by Eqn. 6?
 The easiest way to explore the shift in $\beta$
 due to fermions  is by tracking the confinement-
deconfinement transition
as a function of quark mass and number of flavors.

The quenched phase transition at $N_T=4$ is at $\b_{c}^Q=5.69(1)$
[\cite{SGREVIEW}].
Introducing $N_f$ flavors of fermions with mass $m$ will shift the
transition to $\b_{c}^{N_f}=\b_{c}^Q-\D\b$. If $m $ is such that the
fermionic action can be considered pure gluonic at low energies
 then $\D\b(N_f,m)=N_f \D\b_1(m)$. If, in
addition, $m$ is large enough that the perturbative formula is valid,
$\D\b_1(m)$ is given by Eqn. 6.
Thus we expect the following behavior for the shift $\Delta\beta(N_f,m)$:
For $m >> \Lambda$ where the fermion and gluon mass scales are well
separated we expect to see universal behavior $\Delta\beta/N_f=f(m)$
where $f(m)$ is given by Eqn. 6. For smaller $m$
we  expect Eqn. 6 to fail quantitatively.
However, it might happen that $\Delta\beta/N_f$ is still
some universal function of the quark mass. Finally, when the fermion scale
is the same order as the gauge scale one can no longer replace the fermions
by an effective gauge action. The shift $\Delta\beta/N_f$
would then be different for different $N_f$, $N_T$.
Measuring the finite temperature transition for different $N_f$ and $m$
values makes it possible to distinguish the different scenarios.

The finite temperature transition is
first order for the pure gauge theory, and is stable under the inclusion
of heavy fermions. With decreasing quark mass the location of the transition
shifts downward in $\beta$.
 At some point the deconfinement transition line terminates (at
sufficiently light quark mass). We might still be able to
track the crossover point
as a function of $N_f$ and $m$.
As long as the fermionic spectrum remains heavy compared to the low
energy gluon spectrum, the system  could still
 be described by an effective gauge
action and Eqn. 6 could be  valid.

 At very small or zero quark mass (depending on the number of light flavors)
there is a second transition whose behavior is thought to be primarily
chirally-restoring.
At this transition the role of the fermions is fundamental  and one would
not expect the decoupling of the gluonic and fermionic spectrum.

In a model with $N_l$ light flavors of mass $m_l$
 and $N_h$ heavy flavors of mass $m_h$, we also expect
that the transition should be shifted by the heavy flavors:
$\beta_c(N_l,m_l,N_h,m_h) = \beta_c(N_l,m_l) + \Delta \beta(N_h,m_h)$
where $\Delta \beta$ is given by Eqn. 6. It is a phenomenologically
interesting question to ask, ``How light is still heavy?''
For example, several groups have recently performed simulations with
$N_l=2$ and $N_h=1$ in an attempt to model the deconfinement transition
in the real world of two light ($u,d$) quarks and one strange quark.
To the extent that Eqn. 6 predicts the shift in lattice  critical
coupling, the heavy flavor is merely renormalizing the gauge coupling
and contributing no new physics.

In the above consideration we had to assume the relation $\b=6/g^2_0$ -
the induced gauge coupling is expressed through the bare continuum
coupling $g^2_0$ while in a lattice simulation one uses the coefficient
of the plaquette term $\b$. $\b=6/g^2_0$ should hold in the continuum,
large $\b$ limit; one
expects to encounter deviations when the finite temperature transition
happens in the strong coupling (small $\b$) region.

The $\D\b$ values of Table 1 are calculated on an infinite lattice. On
a finite lattice simulations there are corrections to it. For example,
 at $N_T=4$ the fermions induce an explicit Polyakov loop term in the
action which is the same order as the induced $\beta$ of the gauge coupling
($\kappa^4$ in hopping parameter expansion).
In all simulations with dynamical fermions at $N_T=4$,
we have a nonzero expectation value for the Polyakov loop, even
in the confined phase.
That is an additional source of error in comparing Eqn. 6  to finite
lattice numerical results.

Now we consider a number of cases. We have chosen to focus on staggered
fermions since it is easier to make a connection to Eqn. 6 with them than
with Wilson fermions.
In Sect. 5 we discuss a possible connection to the
confinement-deconfinement transition for Wilson fermions.

In the following we will translate numerical data to express the shift
in the gauge coupling caused by one of the fermions only,
$\D\b_1=(\b^Q_c-\b^{N_f}_c)/N_f$. Here $\b^Q_c$ is the Monte Carlo
quenched critical coupling and $\b^{N_f}_c$ is the Monte Carlo $N_f$
flavor critical coupling.
This way we can compare simulations with
different $N_f$ and $N_T$ values.

\subhead{4.1 $N_f=24$}
Table II  shows the result of a 24 flavor staggered fermion
simulation on $6^3\times4$ lattices
for several mass values.
These simulations were done by us and use a version of a code written by
the MILC collaboration\refto{SG_JULICH}.
 We employ the
hybrid molecular dynamics algorithm described in Ref.~\cite{HMD}.
We have defined the dynamical fermion fields on all sites of the lattice,
so that the ``natural'' number of flavors in the simulation is
a multiple of 8.
Simulations with large $N_f$ require a very small timestep compared to
ones with small $N_f$ since the fermion force in the microcanonical
evolution equations scales linearly in $N_f$. For example for $m=0.5$
stepsize $\D t=0.020$ is needed.
The $N_f=24$ transition is very sharp. Fig. 1 shows the expectation
value of the Polyakov loop at $m=0.5$. At $\b=4.62$ the time evolution
shows tunneling between two states (Fig. 2), the transition is probably
first order.
The data points agree with the analytical
prediction for $m>0.25$.  At $m=0.25$ simulations with $N_f\le8$
agree with the analytic prediction. The deviation here should be
attributed to the fact that $\b_{c}=3.90(5)$
is a very strong coupling where $\b = 6/g^2_0$
does not hold anymore.
\if\preprint Y \psfigure  252 108 {Figure 1} {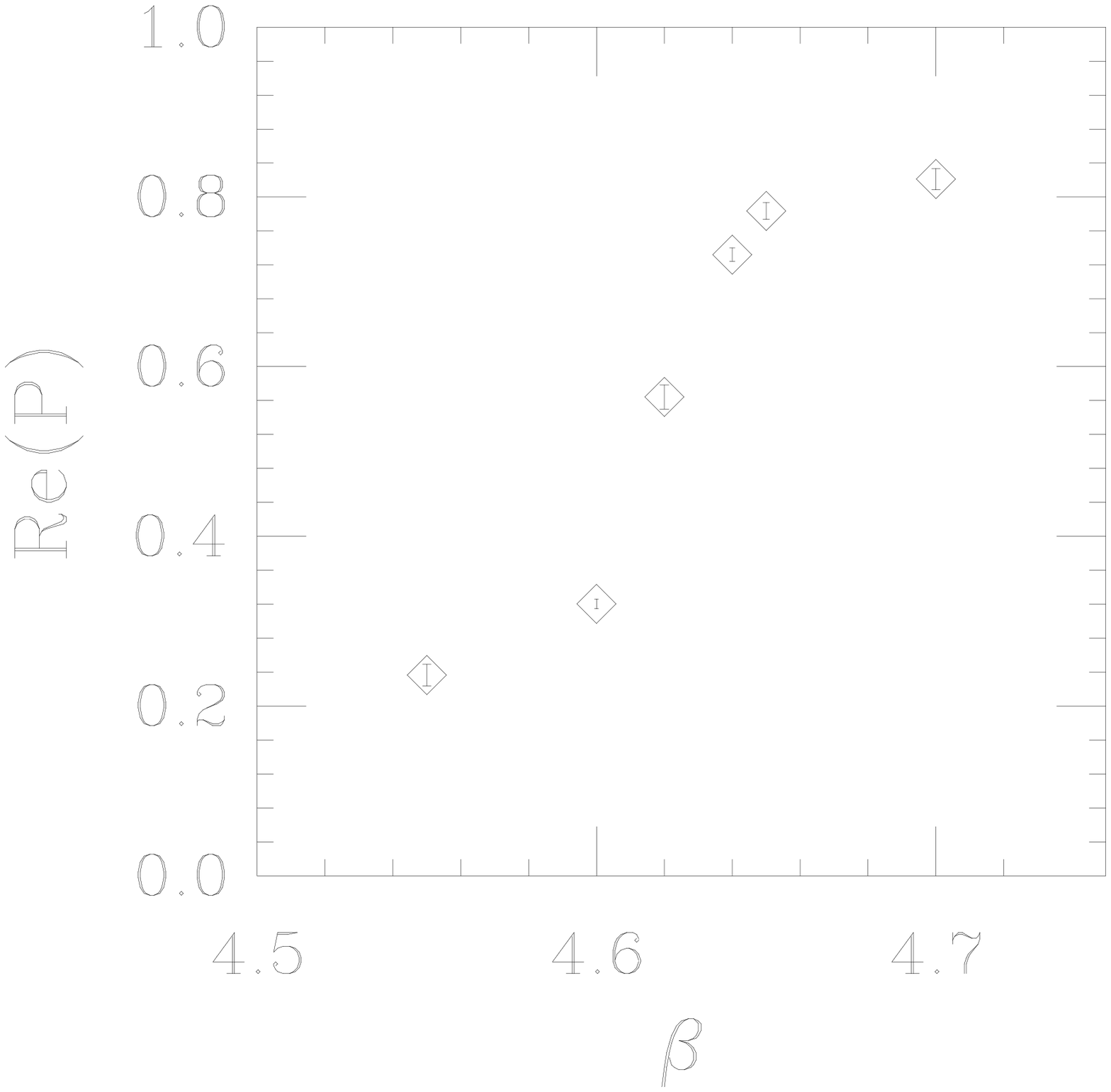} {
The real part of the Polyakov loop for $m=0.5$, $N_f=24$.}\fi

\if\preprint Y \psfigure  252 108 {Figure 2} {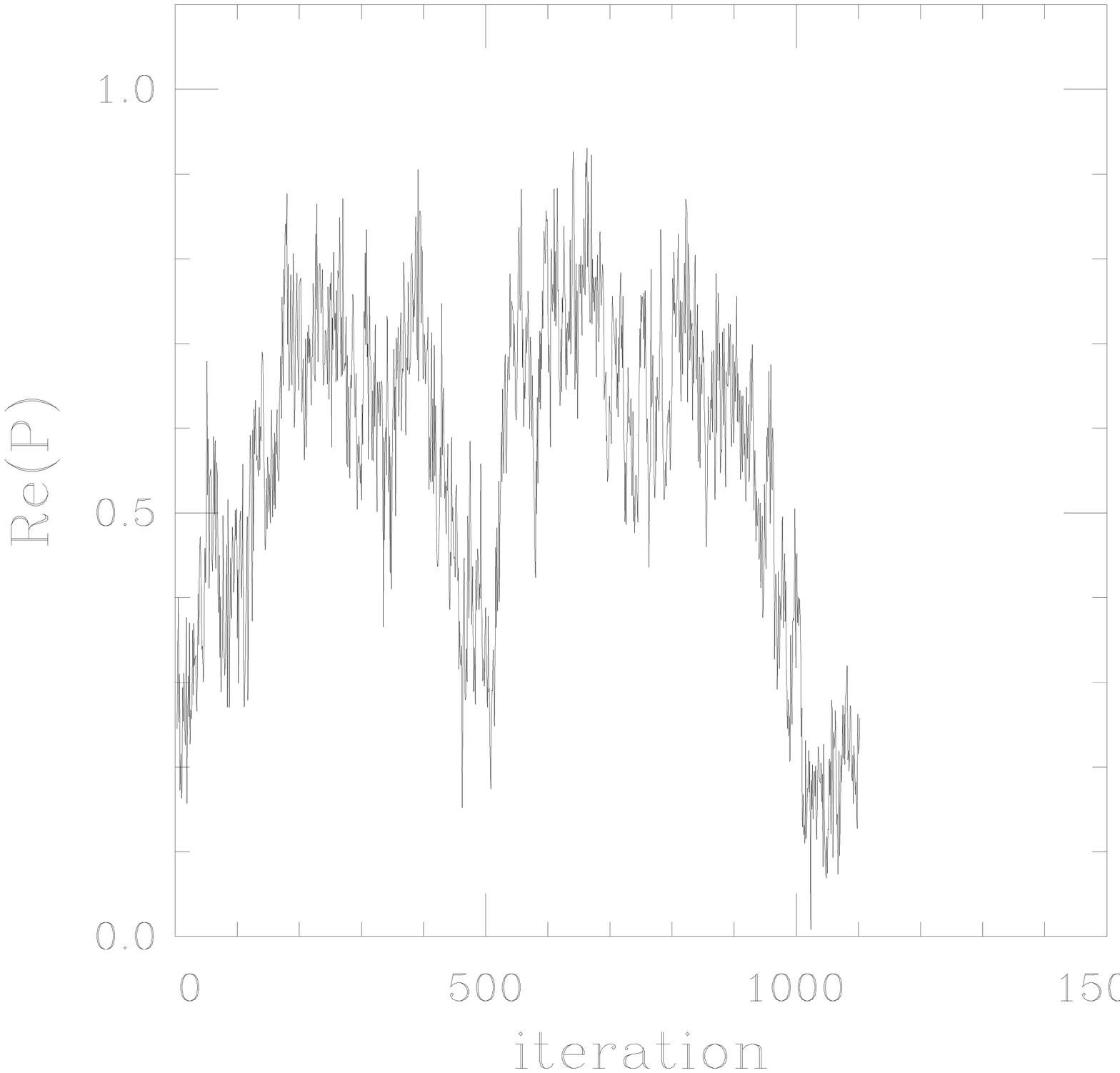} {
Time evolution of the real part of the Polyakov loop for $m=0.5$, $N_f=24$.}\fi

\subhead{4.2 $N_f=17$}
These data  (shown in Table III)
are from runs using the Langevin updating algorithm on
$N_T=4$ lattices\rlap.\refto{COLUM817} The
analytic formula consistently overestimates the shift in $\beta_c$.
This is hard to understand given that the $N_f=24$ and $N_f=8$ simulations
(see below)
are well represented by the formula. However, the Langevin timestep $\Delta
t_L$
is related to the timestep of microcanonical simulations $\Delta t_M$
by\refto{ARGENTINA} $\Delta t_L = (\Delta t_M)^2/2$. These simulations are
performed at
$\Delta t_L=0.01$ corresponding to $\Delta t_M=0.14$, which is known
to be large enough to induce sizable integration timestep errors.

\subhead{4.3 $N_f=8$}
These data, shown in Table IV,
are also from runs using the Langevin updating algorithm on
 $N_T=4$ and 6 lattices.
For smaller mass values the results are very sensitive to the step size
used in the simulations. For too large $\D t$ the transition is
generally overestimated, so the shift $\D\b$ is underestimated.
The analytic formula accurately predicts the location of the transition
or crossover point for the larger  values of the quark mass studied.
For smaller quark mass values the agreement is still reasonable though
$\D\b^{MC}$ is consistently smaller than the analytic prediction.
At very small $m$  and $N_T>8$ a number of
 authors\refto{COLUM8, OTHERWEIRD,fuk2}
have seen a transition which may be a bulk transition.
Our analytic formula does not predict this transition. On the other hand
at such small mass values Eqn. 6 is not expected to be valid anymore.

\subhead{4.4 $N_f=4$}
These simulations, shown in Table V, do not show a phase transition at moderate
values of
the quark mass. At small $m$ they show a first order transition
which is believed to be associated with chiral restoration. The location
of the transition/crossover is well predicted by Eqn. 6 down to
$m=0.05$. For $N_T=4$  at $m=0.073$ the first order chiral
 transition switches on\rlap.\refto{SGREVIEW} It
is surprising that Eqn. 6 is still valid. One can see deviation from the
analytic formula for $m \le 0.025$.

\subhead{4.5 $N_f=2$}
Most of the $N_f=2$ simulations were performed at very light values of
the quark mass. They do not show a phase transition; instead, they show
a smooth crossover from a chirally broken phase to a chirally restored one.
Nevertheless, the location of the crossover point is very well tracked by the
analytic formula, even at very light values of the quark mass.
The results are collected in Table VI.

\subhead{4.6 Summary}
Fig. 3 contains our $N_f=24$
data and  $N_f=17$, 8, 4 and 2 data for $N_T=4-8$. Some of these data
points correspond to real first order transitions, others describe just
a crossover. For larger masses they correspond to the $Z_3$ transition,
for  smaller masses they describe the chiral
transition.
The agreement with the analytic prediction, especially with smaller
$N_f$, is remarkable even for masses as small as $m=0.05$ or below.
The fact that the data appear to lie on a universal curve
 is a signal that the fermions induce an effective $\beta$
whose strength is linear in $N_f$ at fixed quark mass,  down to
very small mass.

One can conclude that the effect of dynamical fermions for finite
temperature transition is no more than an induced effective gauge
coupling even for fairly small  ($m \ge 0.05$)
fermion masses.

\if\preprint Y \psfigure  252 108 {Figure 3} {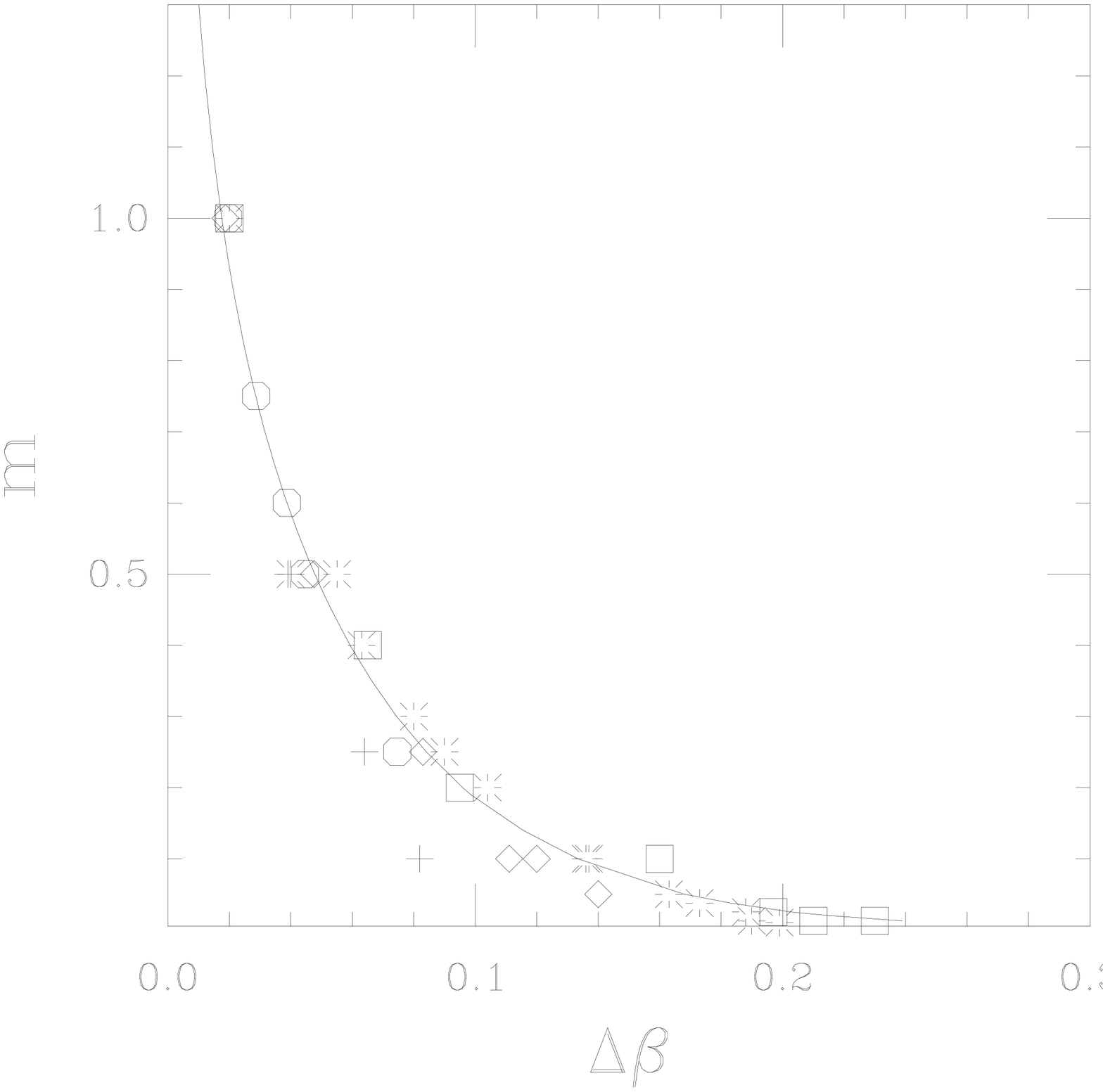} {
The induced gauge coupling divided by the number of flavors,
$\Delta \beta/N_f$,
from the simulations described in this paper, compared with the curve from
Eqn. 6, as a function of quark mass.
Data are labeled with
octagons for $N_f=24$,
pluses for $N_f=17$,
diamonds for $N_f=8$,
bursts for $N_f=4$, and
squares for $N_f=2$.
}\fi

\vfill\supereject

\heading{Light and Heavy Flavors Together}
\subhead{5.1 $N_f=2 + 1$}
Several group studied QCD with two light and one heavy flavors. In these
simulations the heavy fermion mass was between 0.1 and 0.25 and 4 to 20
times heavier than the light species. According to the previous chapter
the effect of fermions with $m=0.1-0.25$ on the finite temperature
transition is well described by an effective action with induced gauge
coupling given by Eqn. 6. Unless there is a strong interaction between the
light and heavy flavors one would expect the same here. This hypothesis
can be tested by comparing the shifts caused by a light species in the
$N_f=2$ and $N_f=2+1$ simulations assuming the effect of the heavy
flavor can be described by Eqn. 6.
Table VII shows this comparison for
recent simulations.  Here we use the notation
$$\eqalign{
&\D\b^{2+1}_l=(\b^Q_c-(\b_c^{2+1}+\D\b_h^{anal}))/2 \cr
&\D\b^{2}_l=(\b^Q_c-\b_c^{2})/2 .
}\eqno{(14)}
$$
where $\b^Q_c$ is the Monte Carlo quenched critical coupling, $\b^{2+1}_c$
is the Monte Carlo $N_f=2+1$ critical coupling and $\b^2_c$ is the Monte
Carlo $N_f=2$ critical coupling. $\D\b^{anal}_h$ is the analytically
predicted induced gauge coupling due to the heavy fermions.
 If $\Delta\beta_l^{2+1} = \Delta\beta_l^{2}$,
 the shift in coupling due to the light quarks is independent
of the presence of the heavy quark.
If Eqn. 6 is valid for the light species, too,
 we expect $\Delta\beta_l^{2+1} = \Delta\beta_l^{2} =
\Delta\beta_l^{anal}$.The shifts from the $N_f=2+1$ and $N_f=2$
simulations agree within errors though $\D\b_l^{2+1}$ is consistently
smaller than $\D\b_l^2$ which agrees with the analytic prediction
 $\D\b^{anal}_l$ .

\subhead{5.2 $N_f=2 + 8$}
As another test of the interplay of light and heavy flavors, we
 performed  simulations with two light flavors and eight heavy flavors
on $8^3 \times 4$ lattices.

We were motivated to perform these studies by consideration of the
deconfinement transition with Wilson fermions.
Wilson fermions have 15 doublers for each light species.
What is the role of the doublers? Do they only generate an effective
gauge coupling or can they influence the low energy spectrum in a
non-trivial way?
 Perturbatively,
the 4 doublers sitting at the nearest edges of the Brillouin zone with
one component of momentum equal to $\pi$ and three components of momentum equal
to zero are much lighter  than the
others. It is plausible to assume that they give the most important
contribution to the effective action, i.e. 2 flavors of Wilson fermions
can be modeled as 2+8 flavors of staggered fermions.
(Wilson fermions and 2+8 flavors of staggered fermions are of course not
identical,  since they have different flavor and chiral symmetry properties.
This approach  just  models the effect of doublers.)

For the heavy flavors we choose mass values $m_h=0.88$, 0.77, 0.665, and 0.4,
corresponding roughly to the bare Wilson doubler masses at $\kappa=0.17-0.21$.
The light masses were chosen as listed in Table VIII.

These runs  were performed on  the
 Intel
iPSC/860 hypercube at the San Diego Supercomputer Center.
 The iPSC/860 and the code are
described briefly elsewhere \refto{SG_JULICH}.
We used a truly hybrid algorithm for these simulations: the eight heavy
flavors were simulated using the $\Phi$ algorithm of Ref. \cite{HMD},
with a random noise term for the fermions which was refreshed at the start
of each microcanonical trajectory. (The fermion fields were defined on all
sites of the lattice to produce eight flavors.) The two light flavors
were simulated using the $R$ algorithm of Ref. \cite{HMD}; the noisy
estimator for their determinant was updated throughout the simulation.
We also performed a two flavor simulation at $m=.04$ for comparison.
We used integration timesteps of $\Delta t=0.1$ for the $m_l=0.2$ and 0.1
simulations. The smaller quark mass simulations were more sensitive to $\Delta
t$ systematics. We used
 $\Delta t=0.05$ away from the transition for all the $m=.04$ simulations
and switched to $\Delta t=0.02$ near the transitions.

We display plots of the Polyakov loop and $\bar \psi \psi$ for the light
quark from our simulations with light quark mass 0.04  in Fig. 4.
The heavy quark masses are 0.4, 0.665 and $\infty$.
The smaller step size points are shown as squares in the figure.
The transition  for the system with two light and eight
heavy flavors appears to be much sharper than the transition for the system
containing only two light flavors. It might be first order. Note that at this
value of the light quark mass the $N_f=2$ transition is a smooth crossover
and the $N_f=8$ system does not have a first order transition for
$m_h\ge 0.25$ either.
\if\preprint Y \psoddfigure  466 466  -9 {Figure 4} {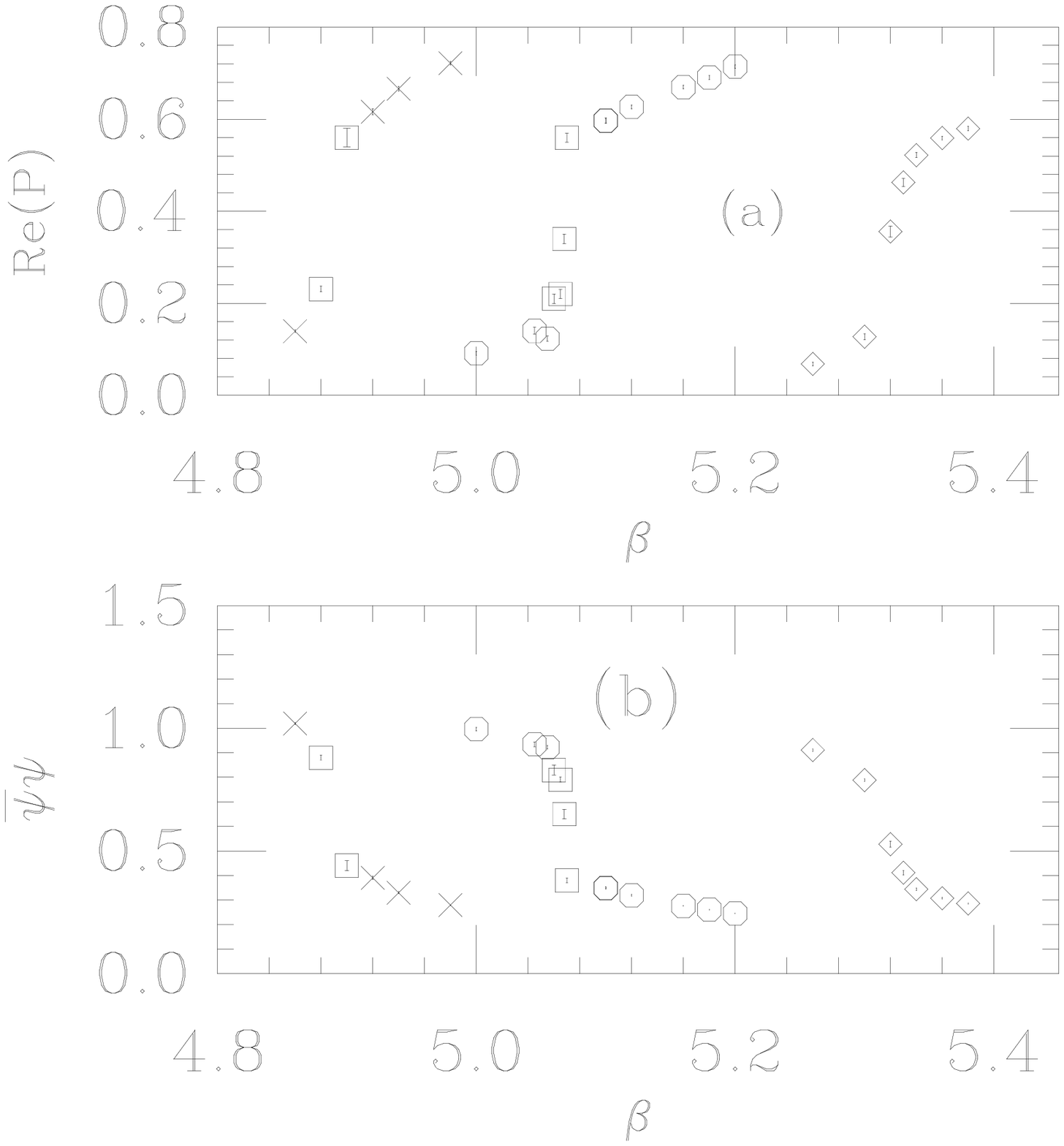} {
Plots of (a) the Polyakov loop and (b) $\bar \psi \psi$ for simulations
with two  flavors of light quarks ($m_l=0.04$) and
either nothing else (diamonds) or eight flavors of heavy quarks, of
mass $m_h=0.665$ (octagons and squares) or 0.4 (crosses and squares).
The squares show data points from simulations with $\Delta t=0.02$;
all other data points used $\Delta t=0.05$.}\fi

In the previous chapter we concluded that the gauge coupling induced by the
heavy flavors is well described by the analytic formula.
Using Eqn. 6 we compute the shift caused by one of the light flavors as
in sect. 5.1 and compare it to the shift observed in the $N_f=2$
simulations.
We present these results in Table VIII.
Our results
for $m_l=0.1$ and 0.2 reproduce the $N_f=2$ simulation results and the
analytic prediction, as we would
expect following the successes recorded in the last chapter. Neither of
our $m_l=0.04$ results agree with the analytic formula. That could be
explained simply as a breakdown of the analytic formula at light quark mass.
However,  with $N_f=2$ or 4, simulations at $m_q=0.05$ still agree
with the analytic formula, as can be seen by comparing Tables V, VI,
and the last entry of Table VII.
What is even more surprising, the $m_h=0.665$ and 0.4 data show a
different shift in $\beta$
from the same light quarks $m_l=0.04$.
These facts, coupled with the qualitative sharpening of the transition
at smaller $m_h$, lead us to conclude that the eight heavy flavors have
an observable influence on the light flavors
in addition to an induced gauge coupling. The assumption that the heavy flavors
are unimportant  at low energies does not seem to hold.

One might expect that this result would be even stronger
if the light fermions are lighter.

One might also expect similar behavior for Wilson fermions. In fact, one
might expect an even stronger effect, since Wilson fermions include
explicit interactions between the light quarks and the  doublers which
are not present in this $2+8$ flavor system.

\vfill\eject
\heading{$\b=0$ limit}

128 flavors of fermions with mass $m\approx 0.4$ induce a gauge coupling
$6/g^2=\b_{ind}\approx 7.6$. That is large enough to deconfine an $N_T=4$
system even when the plaquette gauge coupling is zero. With  large
number of flavors one should see a confining-deconfining pure gauge phase
transition in the $\b=0$ limit as the function of the fermion mass.

The naive analytical prediction in Sect. 2 predicted that for $N_f >16$
flavors the fermions always decouple from the low lying gauge spectrum
even in the $\b=0$ strong coupling limit.
Does that mean that for $N_f>16$ at $\b=0$ one will always find a
deconfining phase transition for some value of the quark mass?

We obviously cannot check this scenario numerically but we can study the
$N_T=4$ finite temperature phase transitions in $m$ at $\b=0$ for different
$N_f$ values. Fig. 5a  shows $m_{crit}$  and 5b shows $\b_{ind}$ at the
phase transition calculated from Eqn. 6 as the function of $N_f$.
Since we observed
strong metastability in all cases, we conclude that
the phase transition with so many fermions is first order.
\if\preprint Y \psoddfigure  466 466 -9 {Figure 5} {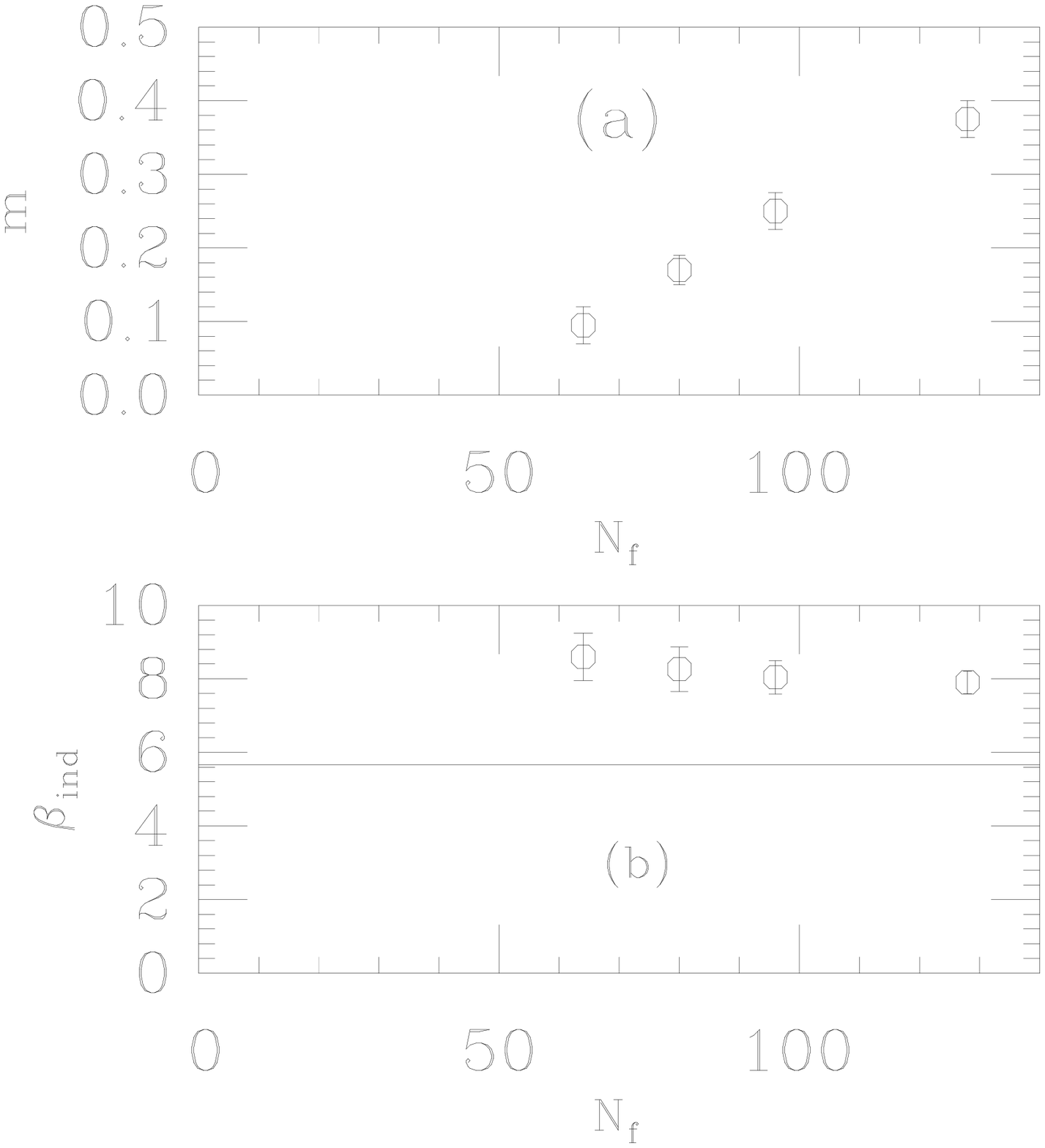} {
(a) Plot of the location of the confinement-deconfinement transition
at $\beta=0$
as a function of quark mass for several values of $N_f$.
(b) The same data, but now interpreted as a function of the induced coupling
inferred from the analytic expression.
The line shows the location of the quenched deconfinement transition.}\fi

The induced $\beta$ lies in the range 7.5 to 8.5 for $N_f\ge 80$. The constancy
of this
result over a wide range of $N_f$ indicates that the fermions do induce
an effective gauge coupling which scales with $N_f$. This
$\b_{ind}$ is not consistent with the quenched critical coupling
$\b^Q_c=5.69$ indicating that $6/g^2 $, the coefficient of
$F_{\mu\nu}F_{\mu\nu}$, does not equal $\b$ for small $\b$ values.

\heading{Conclusion}

We demonstrated that the effects of fermions on the finite temperature
phase transition can be described by an induced effective plaquette term
for masses as low as $m\simeq 0.05$.  The induced coupling is proportional
to the flavor number and is independent of $N_T$. The proportionality
constant is given by a simple 1-loop formula.
It is amazing that the simple formula for a fermion-induced shift
in $\beta$ works so well down to such small quark mass, for degenerate
mass fermions.
 From the point of view of lattice
simulations of QCD, our results show that some
dynamical quarks must be very light
to cause interesting effects. A finite temperature simulation at some
quark mass ought to show an induced $\beta$ which is not given by
the one-loop formula, before one could claim that a $T=0$ simulation
at the same mass would be sensitive to the effects
of dynamical quarks. This is just barely the case in contemporary dynamical
fermion simulations. For example, the  spectroscopy of the HEMCGC simulations
 at $\beta=5.6$ with
$N_f=2$ and $m=0.025$ and 0.01 has been mapped onto quenched simulations at
$\beta=5.935$ and 5.95, respectively\rlap.\refto{HEMCGC90}
 These comparisons correspond to shifts per flavor of $\D\b=.1675$ and
0.175, respectively, to be contrasted with $\Delta\beta^{anal}=0.20$
and 0.25 and finite temperature Monte Carlo shifts of about 0.20 and 0.21.
Thus they are  in a regime where the sea quarks might be important
for long distance dynamics.

We simulated systems with 2 light and 8 heavy flavors to study the
interaction of heavy and light quarks.
For light masses $m_l\ge0.1$ we found no observable effect. For
$m_l=0.04$ interaction with the heavy fermions as heavy as $m=0.665$
can be observed in the finite temperature phase transition. We found the
$N_f=8+2$ transition is very sharp and its location cannot be predicted
from
the $N_f=2$ transition assuming that the effect of the heavy flavors is
 described
by an induced gauge coupling.
These results may have applications to technicolor models, where one
has  to deal with the low energy effects of
large numbers of heavy fermions as well as a small number of light fermions.
Consequences of these results for Wilson fermions remain an open problem.

\vfill\supereject
\subhead{Acknowledgements}
Some of the computations were done on the iPSC/860  at the San
Diego Supercomputer Center.  We are grateful to the SDSC staff and the
personnel of Intel Supercomputing Systems Division for their assistance.
This work was supported by the National Science Foundation under grants
NSF--PHY90--23257,   
NSF--PHY91--01853, 
and by the U.~S. Department of Energy under contract
DE--AC02--91ER--40672. 

\endpage
\if\preprint N
\figurecaptions
\item{1.}
The real part of the Polyakov loop for $m=0.5$, $N_f=24$.

\item{2.}
Time evolution of the real part of the Polyakov loop for $m=0.5$, $N_f=24$.

\item{3.}
The induced gauge coupling divided by the number of flavors,
$\Delta \beta/N_f$,
from the simulations described in this paper, compared with the curve from
Eqn. 6, as a function of quark mass.
Data are labeled with
octagons for $N_f=24$,
pluses for $N_f=17$,
diamonds for $N_f=8$,
bursts for $N_f=4$, and
squares for $N_f=2$.

\item{4.}
Plots of (a) the Polyakov loop and (b) $\bar \psi \psi$ for simulations
with two  flavors of light quarks ($m_l=0.04$) and
either nothing else (diamonds) or eight flavors of heavy quarks, of
mass $m_h=0.665$ (octagons and squares) or 0.4 (crosses and squares).
The squares show data points from simulations with $\Delta t=0.02$;
all other data points used $\Delta t=0.05$.

\item{5.}
(a) Plot of the location of the confinement-deconfinement transition
at $\beta=0$
as a function of quark mass for several values of $N_f$.
(b) The same data, but now interpreted as a function of the induced coupling
inferred from the analytic expression.
The line shows the location of the quenched deconfinement transition.

\endfigurecaptions
\head{Table Captions}

\item{I.}
$\D\b$ as predicted by Eqn. 6 as the function of the quark
mass.

\item{II.}
$N_f=24$ simulations on $6^3 \times 4$ lattices
performed by us. All the phase transitions, except the $m=1.00$ one are
very sharp, probably first order. At $m=1.00$ there is only a broad
crossover around $\b=5.24$.

\item{III.}
$N_f=17$ simulations, from Ref. \cite{COLUM817}, at $N_T=4$.

\item{IV.}
$N_f=8$ simulations, from Ref. \cite{COLUM8}

\item{V.}
$N_f=4$ simulations performed by several groups.

\item{VI.}
$N_f=2$ simulations performed by several groups.

\item{VII.}
$N_f=2+1$ simulations from Ref. \cite{COLUM2+1} and
\cite{KS2+1}.

\item{VIII.}
 $N_f=2+8$ simulations performed by us.

\vfill\eject
\fi

\TABLEcap{I}{ $\D\b$ as predicted by Eqn. 6 as the function of the quark
mass.}
\vskip-24pt
$$
{\vbox{\offinterlineskip\halign{
\vrule\fstrut\quad\hfil#\hfil\quad&&\fstrut\quad\hfil#\hfil\quad\cr
\dbline\notext

\vrule height 18pt depth 7pt width 0pt
$m$ & .025 & .05 & .1 & .2 & .3 & .4 & .5 & .75 & 1.0 \endrule
\sgline
$\D\b$ & .203 & .168 & .133 & .096 & .074 & .059 & .048 & .029 &
.020 \endrule
\sgline
}}
}
$$

\TABLEcap{II}{ $N_f=24$ simulations on $6^3 \times 4$ lattices
performed by us. All the phase transitions, except the $m=1.00$ one are
very sharp, probably first order. At $m=1.00$ there is only a broad
crossover around $\b=5.24$.}
\vskip-24pt
$$
{\vbox{\offinterlineskip\halign{
\vrule\fstrut\quad\hfil#\hfil\quad&&\fstrut\quad\hfil#\hfil\quad\cr
\dbline\notext

\vrule height 18pt depth 7pt width 0pt
$m$ & $\beta_c$ & $\Delta \beta^{MC}/N_f$ &  $\Delta \beta^{anal}/N_f$\endrule
\sgline
1.00  &  5.24(4)  &  .0188(16) & .0175 \endrule
0.75  &  5.00(2)  &  .0287(8) & .0286 \endrule
0.60  &  4.76(2)  &  .0387(8) & .0388 \endrule
0.50  &  4.62(2)  &  .0446(8) & .0478 \endrule
0.25  &  3.90(5)  &  .0746(20) & .0840 \endrule
\sgline
}}
}
$$

\TABLEcap{III}{ $N_f=17$ simulations, from Ref. \cite{COLUM817}, at $N_T=4$.}
\vskip-24pt
$$
{\vbox{\offinterlineskip\halign{
\vrule\fstrut\quad\hfil#\hfil\quad&&\fstrut\quad\hfil#\hfil\quad\cr
\dbline\notext

\vrule height 18pt depth 7pt width 0pt
$m$ & $\beta_c$ & $\Delta \beta^{MC}/N_f$ &  $\Delta \beta^{anal}/N_f$\endrule
\sgline
0.50  &  5.025  &  .039 & .0478 \endrule
0.25  &  4.6(1) & .064 & .084 \endrule
0.10  &  4.3(1)  & .082 & .133 \endrule
\sgline
}}
}
$$

\TABLEcap{IV}{ $N_f=8$ simulations, from Ref. \cite{COLUM8}}
\vskip-24pt
$$
{\vbox{\offinterlineskip\halign{
\vrule\fstrut\quad\hfil#\hfil\quad&&\fstrut\quad\hfil#\hfil\quad\cr
\dbline\notext

\vrule height 18pt depth 7pt width 0pt
$m$ & $\beta_c$ & $\Delta \beta^{MC}/N_f$ &  $\Delta \beta^{anal}/N_f$
& $N_T$\endrule
\sgline
1.0   &  5.54  &   .0187 & .0175 & 4 \endrule
0.50  &  5.31  &  .0475 & .0478 & 4 \endrule
0.25  &  5.025 & .083 & .084 & 4 \endrule
0.10  &  4.80(1) & .111(1) & .133 & 4 \endrule
0.10  &  4.95  &   .12(1) & .133 & 6 \endrule
0.05  &  4.75  &   .14(1) & .168 & 6 \endrule
\sgline
}}
}
$$

\vfill\eject

\TABLEcap{V}{ $N_f=4$ simulations
performed by several groups.}
\vskip-24pt
$$
{\vbox{\offinterlineskip\halign{
\vrule\fstrut\quad\hfil#\hfil\quad&&\fstrut\quad\hfil#\hfil\quad\cr
\dbline\notext

\vrule height 18pt depth 7pt width 0pt
$m$ & $\beta_c$ & $\Delta \beta^{MC}/N_f$ &  $\Delta \beta^{anal}/N_f$
& $N_T$ & Ref.\endrule
\sgline
0.5  & 5.50 & .04 & .048 & 4 & \cite{fuk2} \endrule
0.5 & 5.45 & .055 & .048 & 4 & \cite{Gott_nf24} \endrule
0.4 & 5.42 & .063 & .059 & 4 & \cite{Gott_nf24} \endrule
0.3 & 5.35 & .08 & .074 & 4 & \cite{fuk2} \endrule
0.2  &  5.255(5)  &  .104(2) & .0958 & 4 &  \cite{Gott_nf4}\endrule
0.1  &  5.130(5)  &  .136(2) & .1334 &4 &  \cite{Gott_nf4}\endrule
0.05 & 5.04 & .163 & .168 & 4 & \cite{COLUM4} \endrule
0.0375 & 4.99 & .173 & .183 & 4 & \cite{KS2} \endrule
0.0375 & 5.02  & .168& .183 & 4 & \cite{COLUM4} \endrule
0.025  &  4.98(2)  &  .175 & .203  &4 &  \cite{Gott_nf4}\endrule
0.0125 & 4.919 & .190 & .240 & 4 & \cite{KS2} \endrule
0.01 & 4.95 & .185 & .25 & 4 & \cite{COLUM4} \endrule
0.25  & 5.509 & .090 & .084 & 6 & \cite{KS1} \endrule
0.10 & 5.322 & .137 & .133 & 6 & \cite{KS1} \endrule
0.075 & 5.25 & .155 & .149 & 6 & \cite{KS1} \endrule
0.065 & 5.22 & .162 & .155 & 6 & \cite{KS1} \endrule
0.05 & 5.175  & .174 & .168 & 6 & \cite{KS1} \endrule
0.025 & 5.130 & .187 & .203 & 6 & \cite{COLUM4} \endrule
0.01 & 5.08 & .199& .25 & 6 & \cite{COLUM4} \endrule
0.025 & 5.25 & .188 & .203 & 8 & \cite{MTc} \endrule
0.01 & 5.15(5) & .213& .25 & 8 & \cite{MTc8} \endrule
\sgline
}}
}
$$
\vfill\eject

\TABLEcap{VI}{ $N_f=2$ simulations
performed by several groups.}
\vskip-24pt
$$
{\vbox{\offinterlineskip\halign{
\vrule\fstrut\quad\hfil#\hfil\quad&&\fstrut\quad\hfil#\hfil\quad\cr
\dbline\notext

\vrule height 18pt depth 7pt width 0pt
$m$ & $\beta_c$ & $\Delta \beta^{MC}/N_f$ &  $\Delta \beta^{anal}/N_f$
& $N_T$ & Ref.\endrule
\sgline
1.0 & 5.63 & .02 & .018 & 4 & \cite{fuk2} \endrule
0.4 & 5.54 & .065 & .059 & 4 & \cite{fuk2} \endrule
0.2 & 5.48 & .095 & .096 & 4 & \cite{fuk2} \endrule
0.1  &  5.38  &  .15 & .133 & 4 &  \cite{Gott_nf24}\endrule
0.05 & 5.34 & .165 & .168 & 4 & \cite{fuk2} \endrule
0.025  &  5.2875  &  .197 & .203  & 4 &  \cite{Gott_nf24}\endrule
0.0125 & 5.271 & .21 & .24 & 4 & \cite{fuk1} \endrule
0.01 & 5.265(10) & .212 & .25 & 4 & \cite{COLUM2+1} \endrule
0.025 & 5.445 & .212 & .203 & 6 & \cite{nt6} \endrule
0.0125 & 5.42(1) & .225(10) & .239 & 6 & \cite{nt6} \endrule
0.0125 & 5.5375 & .23 & .239 & 8 & \cite{nt8} \endrule
\sgline
}}
}
$$

\vfill\eject

\TABLEcap{VII}{ $N_f=2+1$ simulations from Ref. \cite{COLUM2+1} and
\cite{KS2+1}.
}
\vskip-24pt
$$
{\vbox{\offinterlineskip\halign{
\vrule\fstrut\quad\hfil#\hfil\quad&&\fstrut\quad\hfil#\hfil\quad\cr
\dbline\notext

\vrule height 18pt depth 7pt width 0pt
$m_l$ & $m_h$ & $\beta_c^{2+1}$ &$\D\b_h^{anal}$ & $\D\b_l^{2+1}$ &
$\Delta \beta^{2}_l$ & $\Delta \beta^{anal}_{l}$ &  $N_T$ \endrule
\sgline
0.025 & 0.025 & 5.132(2) & .20 & .18 & .20 & .20 & 4 \endrule
0.025 & 0.10 & 5.171 & .13 & .20 & .20 & .20 & 4 \endrule
0.0125 & 0.25 & 5.199(2) & .084 & .20 & .23 & .24 & 4 \endrule
0.00833 & 0.1667 & 5.325(25) & .11 & .22 & - & .26 &  6 \endrule
\sgline
}}
}
$$

\TABLEcap{VIII}{ $N_f=2+8$ simulations
performed by us. }
\vskip-24pt
$$
{\vbox{\offinterlineskip\halign{
\vrule\fstrut\quad\hfil#\hfil\quad&&\fstrut\quad\hfil#\hfil\quad\cr
\dbline\notext

\vrule height 18pt depth 7pt width 0pt
$m_l$ & $m_h$ & $\beta_c^{8+2}$ & $\Delta\beta_h^{anal}$
  &$ \Delta\beta^{2+8}_l$ &  $\D\b^2_l$ &
$\Delta \beta^{anal}_l$  \endrule
\sgline
0.04 & 0.665 & 5.065(5) & .271 & .170(3) & .183(5) & .18 \endrule
0.04 & 0.40 & 4.89(1) & .474 & .155(5) & .183(5) & .18  \endrule
0.10 & 0.77 & 5.18(1) & .22 & .138(5)  & .15 & .133 \endrule
0.20 & 0.88 & 5.275(25) & .18 & .11(2) & .095 & .096 \endrule
0.04 & $\infty$ & 5.275(25) & 0 & - & .183(5) & .18 \endrule
\sgline
}}
}
$$

\references

\refis{HH}
A. Hasenfratz and P. Hasenfratz,  \journal Phys. Lett.,
B297, 166, 1992.

\refis{L92}
A. Hasenfratz, to appear in the Proceedings of Lattice 92.

\refis{MK}
 V.A. Kazakov,  A.A. Migdal, preprint  PUPT-1322, June 1992.

\refis{KS2+1}
 J. B.~Kogut, D. K.~Sinclair
and K.C.~Wang, \journal Phys. Lett., B263, 101, 1991.

\refis{COLUM2+1}
F.R.~Brown et al., \prl 65, 2491, 1990.

\refis{Gott_nf4}
S.~Gottlieb at al, \journal Phys. Rev., D40, 2389, 1989.

\refis{Gott_nf24}
S.~Gottlieb, W.~Liu, D.~Toussaint, R.~L.~Renken, R.~L.~Sugar,
\journal Phys. Rev., D35, 3972, 1987.

\refis{OTHERWEIRD}
J.~B.~Kogut, J.~Polonyi, H.~W.~Wyld, and D.~K.~Sinclair, \prl 54, 1475, 1985;
J.~B.~Kogut and D.~K.~Sinclair, \journal Nucl. Phys,  B295[FS21], 465, 1988;

\refis{SGREVIEW} For a recent review of the present status of finite
temperature QCD simulations see S.~Gottlieb, \journal Nucl. Physics B
(Proc. Suppl.), 20, 247, 1991.

\refis{HMD}
S.~Gottlieb, W.~Liu, D.~Toussaint, R.L.~Renken and R.L.~Sugar,
\prd 35, 2531, 1987.

\refis{COLUM817}
S.~Ohta, S.~Kim, \journal Phys. Rev., D44, 504, 1991;
and  \journal Phys. Rev., D46, 3607, 1992.

\refis{COLUM8}
F.~R.~Brown, H.~Chen, N.~H.~Christ, Z.~Dong, R.~D.~Mawhinney, W.~Schaffer,
A.~Vaccarino,
\journal Phys. Rev., D46, 5655, 1992.

\refis{KS1}
J.~B.~Kogut and D.~K.~Sinclair  \journal Nucl. Phys., B280[FS18], 625, 1986.

\refis{KS2}
J.~B.~Kogut at al,  \journal Nucl. Phys., B290[FS20], 431, 1987.

\refis{nt6}
C.~Bernard, M.~C.~Ogilvie,
T.~A.~DeGrand,C.~DeTar,S.~Gottlieb,A.~Krasnitz, R.~L.~ Sugar,
D.~Toussaint, \journal Phys. Rev., D45, 3854, 1992.

\refis{fuk1}
M.~Fukugita, H.~Mino,
M.~Okawa, A.~Ukawa
\journal Phys. Rev., D42, 2936, 1990.

\refis{COLUM4}
F.~R.~Brown, F~.P.~Butler, H.~Chen, N.~H.~Christ,
Z.~Dong, W.~Schaffer, L.~I.~Unger, A.~Vaccarino
\journal Phys. Lett., B251, 181, 1990.

\refis{MTc8} %
MT(c) Collaboration (R.~V.~Gavai, et al.)
\journal Phys. Lett., B241, 567, 1990.

\refis{MTc} %
MT(c) Collaboration (R.~V.~Gavai, et al.)
\journal Phys. Lett., B232, 491, 1989.

\refis{nt8} %
S.~Gottlieb at al,  LATTICE '91 Conf., Tsukuba,
\journal Nucl. Phys., B26 (Proc. Suppl.), 308, 1992.

\refis{fuk2}
M.~Fukugita, S.~Ohta, A.~Ukawa,
\journal Phys. Rev. Lett., 60, 178, 1988.

\refis{SG_JULICH} ``QCD on the iPSC/860'',
C.~Bernard, T.~DeGrand, C.~DeTar, S.~Gottlieb, A.~Krasnitz,
M.C.~Ogilvie, R.L.~Sugar, and D.~Toussaint,
in {\it Workshop on Fermion Algorithms}, edited by H.~J.~Hermann and
F.~Karsch, (World Scientific, Singapore, 1991).

\refis{ARGENTINA}
See D. Toussaint, \journal Comp. Phys. Comm., 56, 69, 1989.

\refis{HEMCGC90}
K. M. Bitar, et.~al., \journal Nucl. Phys., B (Proc. Suppl.) 20, 362, 1991.

\endreferences
\endit